\documentclass[%
preprint,
superscriptaddress,
amsmath,amssymb,
prl,
]{revtex4-1}

\usepackage[dvipdfmx]{graphicx}
\usepackage{dcolumn}
\usepackage{bm}
\usepackage{braket}
\usepackage{color}

\begin{document}


\title{Anisotropic spin distribution and perpendicular magnetic anisotropy in the layered ferromagnetic semiconductor (Ba,K)(Zn,Mn)$_{2}$As$_{2}$}

\author{Shoya Sakamoto}
\email{Correspondence: shoya.sakamoto@issp.u-tokyo.ac.jp}
\affiliation{Department of Physics, The University of Tokyo, Bunkyo-ku, Tokyo 113-0033, Japan}
\affiliation{The Institute for Solid State Physics, The University of Tokyo, Chiba 277-8581, Japan}
 
\author{Guoqiang Zhao}%
\affiliation{Beijing National Laboratory for Condensed Matter Physics, Institute of Physics, Chinese Academy of Sciences, Beijing 100190, China}%


\author{Goro Shibata}
\affiliation{Department of Physics, The University of Tokyo, Bunkyo-ku, Tokyo 113-0033, Japan}

\author{Zheng Deng}%
\affiliation{Beijing National Laboratory for Condensed Matter Physics, Institute of Physics, Chinese Academy of Sciences, Beijing 100190, China}%

\author{Kan Zhao}%
\affiliation{Beijing National Laboratory for Condensed Matter Physics, Institute of Physics, Chinese Academy of Sciences, Beijing 100190, China}%

\author{Xiancheng Wang}%
\affiliation{Beijing National Laboratory for Condensed Matter Physics, Institute of Physics, Chinese Academy of Sciences, Beijing 100190, China}%

\author{Yosuke Nonaka}
\affiliation{Department of Physics, The University of Tokyo, Bunkyo-ku, Tokyo 113-0033, Japan}

\author{Keisuke Ikeda}
\affiliation{Department of Physics, The University of Tokyo, Bunkyo-ku, Tokyo 113-0033, Japan}

\author{Zhendong Chi}
\affiliation{Department of Physics, The University of Tokyo, Bunkyo-ku, Tokyo 113-0033, Japan}

\author{Yuxuan Wan}
\affiliation{Department of Physics, The University of Tokyo, Bunkyo-ku, Tokyo 113-0033, Japan}

\author{Masahiro Suzuki}
\affiliation{Department of Physics, The University of Tokyo, Bunkyo-ku, Tokyo 113-0033, Japan}

\author{Tsuneharu Koide}
\affiliation{Photon Factory, Institute of Materials Structure Science, High Energy Accelerator Research Organization (KEK), Tsukuba, Ibaraki 305-0801, Japan}

\author{Arata Tanaka}
\affiliation{Department of Quantum Matter, Graduate School of Advanced Sciences of Matter (ADSM), Hiroshima University, Higashi-Hiroshima 739-8530, Japan}

\author{Sadamichi Maekawa}
\affiliation{RIKEN Center for Emergent Matter Science (CEMS), Wako, Saitama 351-0198, Japan}
\affiliation{Kavli Institute for Theoretical Sciences, University of Chinese Academy of Sciences, Beijing,100190, China.}

\author{Yasutomo J. Uemura}
\affiliation{Department of Physics, Columbia University, New York, New York 10027, USA}

\author{Changqing Jin}%
\affiliation{Beijing National Laboratory for Condensed Matter Physics, Institute of Physics, Chinese Academy of Sciences, Beijing 100190, China}%

\author{Atsushi Fujimori}
\affiliation{Department of Physics, The University of Tokyo, Bunkyo-ku, Tokyo 113-0033, Japan}
\affiliation{Department of Applied Physics, Waseda University, Shinjuku-ku, Tokyo, 169-8555, Japan}

\date{\today}

\begin{abstract}
Perpendicular magnetic anisotropy of the new ferromagnetic semiconductor (Ba,K)(Zn,Mn)$_{2}$As$_{2}$ is studied by angle-dependent x-ray magnetic circular dichroism measurements. 
The large magnetic anisotropy with the anisotropy field of 0.85 T is deduced by fitting the Stoner-Wohlfarth model to the magnetic-field-angle dependence of the projected magnetic moment. 
Transverse XMCD spectra highlights the anisotropic distribution of Mn 3$d$ electrons, where the $d_{xz}$ and $d_{yz}$ orbitals are less populated than the $d_{xy}$ state because of the $D_{2d}$ splitting arising from the elongated MnAs$_{4}$ tetrahedra. It is suggested that the magnetic anisotropy originates from the degeneracy lifting of $p$-$d_{xz}$, $d_{yz}$ hybridized states at the Fermi level and resulting energy gain due to spin-orbit coupling when spins are aligned along the $z$ direction.
\end{abstract}

\pacs{Valid PACS appear here}
\maketitle

\begin{figure}
\begin{center}
\includegraphics[width=8.0 cm]{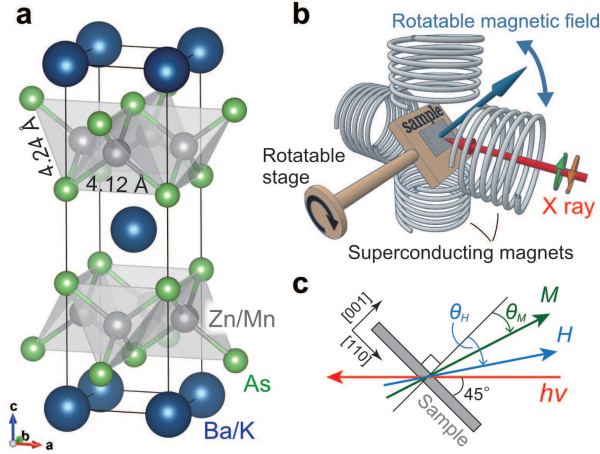}
\caption{Crystal structure of (Ba,K)(Zn,Mn)$_{2}$As$_{2}$ and experimental setup. {\bf a} Unit cell of (Ba,K)(Zn,Mn)$_{2}$As$_{2}$. (Zn/Mn)As$_{4}$ tetrahedron elongated along the $c$-axis by $\sim$6\%. The structures were drawn using a VESTA program \cite{Momma:2011aa}. {\bf b} Schematic figure of the experimental apparatus. {\bf c} Measurement geometry. The sample was placed so that the x-ray incident angle with respect to the sample surface became 45 degrees. $H$ and $M$ denote the magnetic field and the magnetization, and $\theta_H$ ($\theta_M$) denotes the angle of $H$ ($M$) with respect to the sample normal.}
\label{Crystal}
\end{center}
\end{figure}

Ferromagnetic semiconductors (FMSs) have attracted much attention since the discovery of ferromagnetism in (Ga,Mn)As and (In,Mn)As \cite{Ohno:1998aa,Munekata:1989aa,Dietl:2014aa,Jungwirth:2014aa} as they are promising materials for future spintronics applications. 
Recently, a new FMS Ba$_{1-x}$K$_{x}$(Zn$_{1-y}$Mn$_{y}$)$_{2}$As$_{2}$ was synthesized in bulk form \cite{Zhao:2013aa,Zhao:2017aa}, which crystallizes in the tetragonal ThCr$_{2}$Si$_{2}$ structure ($I4$/$mmm$) and is isostructural to 122-type Fe-based superconductors, as shown in Fig. \ref{Crystal}{\bf a}. The host compound BaZn$_{2}$As$_{2}$ is a semiconductor with a narrow band gap of 0.2 eV \cite{Xiao:2014ab}. In this system, one can control the numbers of carriers and spins independently by the heterovalent substitution of K$^{+}$ for Ba$^{2+}$ and the isovalent substitution of Mn$^{2+}$ for Zn$^{2+}$, respectively. Furthermore, with 30\% of K and 15\% of Mn substitution, the Curie temperature ($T_{\rm C}$) reaches 230 K \cite{Zhao:2014aa}, which is higher than $T_{\rm C}$ = 200 K of (Ga,Mn)As \cite{Chen:2011aa}. 
The transport and magnetic properties can also be controlled by external pressure \cite{Sun:2017aa, Zhao:2018aa}. The ferromagnetism is most likely carrier-induced as evidenced by previous experimental and theoretical studies \cite{Zhao:2013aa, Suzuki:2015aa, Suzuki:2015ab,Sun:2016aa,Sun:2017aa, Glasbrenner:2014aa, Yang:2015aa}.


Because the crystal structure is inherently anisotropic, that is, the Ba ions are located between the quasi-two-dimensional (Zn/Mn)As layers and the (Zn/Mn)As$_{4}$ tetrahedra are elongated to the $c$-axis by $\sim$6\% (see Fig. \ref{Crystal}{\bf a}), sizable magnetic anisotropy would be expected. 
In fact, large perpendicular magnetic anisotropy (PMA), where the magnetic easy axis is along the $c$-axis, was observed by SQUID measurements \cite{Wang:2017aa, Zhao:2017aa}, which is useful for future magnetic-memory applications.
In general, magneto-crystalline anisotropy would not appear from the Mn$^{2+}$ high-spin state ($^{6}A_{1}$) because of the lack of orbital magnetic moment.
In the case of (Ga,Mn)As, however, it was reported that biaxial strain from substrate induces perpendicular or in-plane magnetic anisotropy (PMA or IMA) \cite{Abolfath:2001aa, Dietl:2001aa, Liu:2003aa, Sawicki:2004aa, Zemen:2009aa}. This was ascribed to the orbital magnetic moment carried by the holes in the valence bands, which are magnetically coupled with the 3$d$ electrons through $p$-$d$ exchange interaction. 
In the case of (Ba,K)(Zn,Mn)$_{2}$As$_{2}$, however, the valence band top consists of only As 4$p_{z}$ orbital \cite{Xiao:2014ab}, and the system does not have orbital degrees of freedom. Therefore, the orbital magnetic moment of holes alone in host valence bands cannot be responsible for the magnetic anisotropy of (Ba,K)(Zn,Mn)$_{2}$As$_{2}$. 



X-ray magnetic circular dichroism (XMCD) is a powerful method to study magnetic anisotropy because one can directly probe the anisotropy of the spin ($m_{s}$) and orbital ($m_{l}$) magnetic moments. Besides, one can deduce the anisotropic spatial distribution of 3$d$ spins, which appears as the magnetic dipole term ($m_T$) in the XMCD sum rule [21-24], through angle-dependent XMCD (AD-XMCD) measurements.
In particular, XMCD spectra taken under the transverse XMCD (TXMCD) geometry, where the applied magnetic field induces spin magnetic moments perpendicular to the incident x ray, are known to be sensitive to the anisotropic distribution of 3$d$ spins \cite{Laan:2010ab, Shibata:2018aa} because the usually dominant spin contribution to the XMCD spectra vanishes. 
It is worth mentioning that there have been few experimental reports on the observation of TXMCD \cite{Laan:2010ab, Mamiya:2006ab, Shibata:2018aa} because the direction of the magnetic field is usually fixed parallel to the incident x rays in most XMCD measurement systems.

In the present study, we perform AD-XMCD measurements using our custom-designed apparatus and reveal that the large perpendicular magnetic anisotropy of (Ba,K)(Zn,Mn)$_{2}$As$_{2}$ originates from the degeneracy lifting of $p$-$d_{xz}$ and $p$-$d_{yz}$ hybridized orbitals due to spin-orbit interaction and resulting energy gain when spins align to the $z$-direction. 


Ba$_{0.904}$K$_{0.096}$(Zn$_{0.805}$Mn$_{0.195}$)$_{2}$As$_{2}$ single crystals with $T_{\rm C}$ = 60 K were grown by the flux technique (see supplementary material \cite{Suppl_BKZMA} for further details).
AD-XMCD measurements were performed at BL-16A2 of Photon Factory, KEK, where we installed our custom-designed apparatus \cite{Furuse:2013aa,Shibata:2018aa} equipped with two pairs of superconducting magnets so that magnetic fields up to 1 T can be applied to any direction between the incident x-ray direction and a direction perpendicular to it (see Fig. \ref{Crystal}{\bf b} for a schematic drawing). Prior to the measurements, we cleaved the samples {\it in situ} to obtain clean surfaces. Absorption signals were collected in the total-electron-yield mode.
The measurement geometry is shown in Fig.  \ref{Crystal}{\bf c}.
The sample was placed so that the angle between the incident x ray and the $[110]$ direction was 45 degrees. Because the direction of the incident x rays was fixed in the present XMCD measurements, any artifact arising from the saturation effect \cite{Nakajima:1999aa} in the total electron yield mode was ruled out.
XMCD spectra are obtained as the difference between two absorption spectra taken with right- and left-circularly polarized x rays, while x-ray absorption spectroscopy (XAS) spectra as their summation.

In order to extract more information from the experimental XAS and XMCD spectra, we have performed CI cluster-model calculation \cite{Tanaka:1994aa}. 
In the calculation, we assume a tetrahedral [MnAs$_{4}$]$^{-9}$ cluster (Mn$^{3+}$ cluster). We adopt basically the same parameters as those used for (Ga,Mn)As \cite{Kobayashi:2016aa} except that we add a finite $D_{2d}$ splitting that makes the $d_{xz}$ and $d_{yz}$ orbitals lie higher in energy by 0.2 eV than the $d_{xy}$ orbital, and the $d_{x^2-y^2}$ orbital higher in energy by 0.2 eV than $d_{z^2}$. The parameters for the $D_{2d}$ splitting are chosen based on the DFT calculation \cite{Gu:2016aa} which shows a $\sim$0.2 eV splitting for the relevant orbitals (See supplementary material \cite{Suppl_BKZMA} for further details). 
\color{black}



\begin{figure}
\begin{center}
\includegraphics[width=8.0 cm]{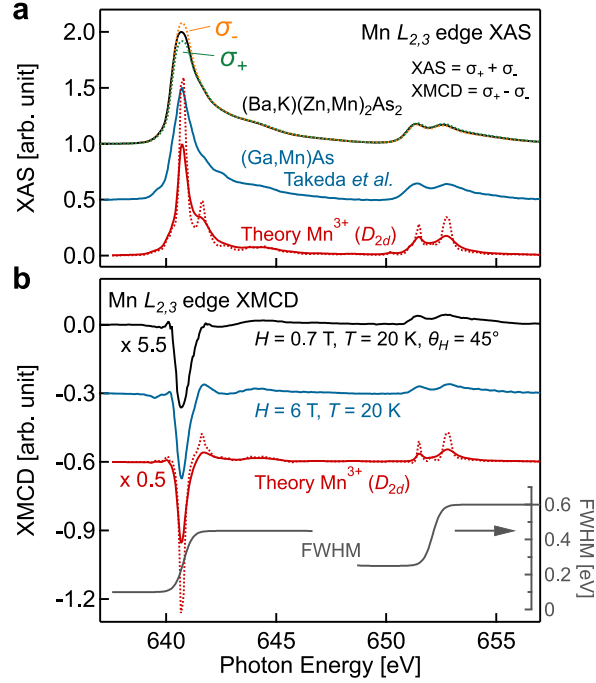}
\caption{Mn $L_{2,3}$-edge XAS and LXMCD spectra of Ba$_{0.904}$K$_{0.096}$(Zn$_{0.805}$Mn$_{0.195}$)$_{2}$As$_{2}$ shown by black curves. The spectra of (Ga,Mn)As \cite{Takeda:2008aa} and the cluster-model calculation are also shown by blue and red dashed curves, respectively. The calculated spectra were broadened (red solid curves) using Lorentzian function with the varying FWHM plotted at the bottom of panel \bf{b}.}
\label{XMCD}
\end{center}
\end{figure}

Figures \ref{XMCD} show the XAS and XMCD spectra of Ba$_{0.904}$K$_{0.096}$(Zn$_{0.805}$Mn$_{0.195}$)$_{2}$As$_{2}$ recorded at the Mn $L_{2,3}$ absorption edges. 
Here, the XMCD spectrum was taken with the magnetic field along the light direction and is dominated by the spin component. We thus refer to this spectrum as the longitudinal XMCD (LXMCD) spectrum hereafter.
The XAS and LXMCD spectra exhibit multiplet features and are very similar to those of (Ga,Mn)As \cite{Takeda:2008aa,Edmonds:2005aa}, which are shown by blue curves in Fig. \ref{XMCD}.
This indicates the localized nature of the Mn 3$d$ electrons being consistent with the carrier-induced ferromagnetism picture, where itinerant holes mediate ferromagnetic interaction between the localized Mn spins. 
Here, the positive peak in the LXMCD spectrum at 642 eV located just above the dominant negative peak at 640 eV is smaller than that of (Ga,Mn)As. 
This may reflect the difference in the electronic structure between (Ga,Mn)As and (Ba,K)(Zn,Mn)$_{2}$As$_{2}$.

The calculated spectra are shown by red dashed curves in Fig. \ref{XMCD}.
The calculated spectra are broadened by a Lorentzian function with a varying full width at half maximum (FWHM) that increases at the $L_{2,3}$ edges in order to reproduce the asymmetric line-shape broadening.
The broadened spectra are shown by red solid curves, and the employed FWHM is plotted at the bottom of Fig \ref{XMCD}{\bf b}. The calculated spectra agree well with the experimental spectra. 
Note that how one broadens the spectra does not change the following discussion and the conclusion.


\begin{figure*}
\begin{center}
\includegraphics[width=17.0 cm]{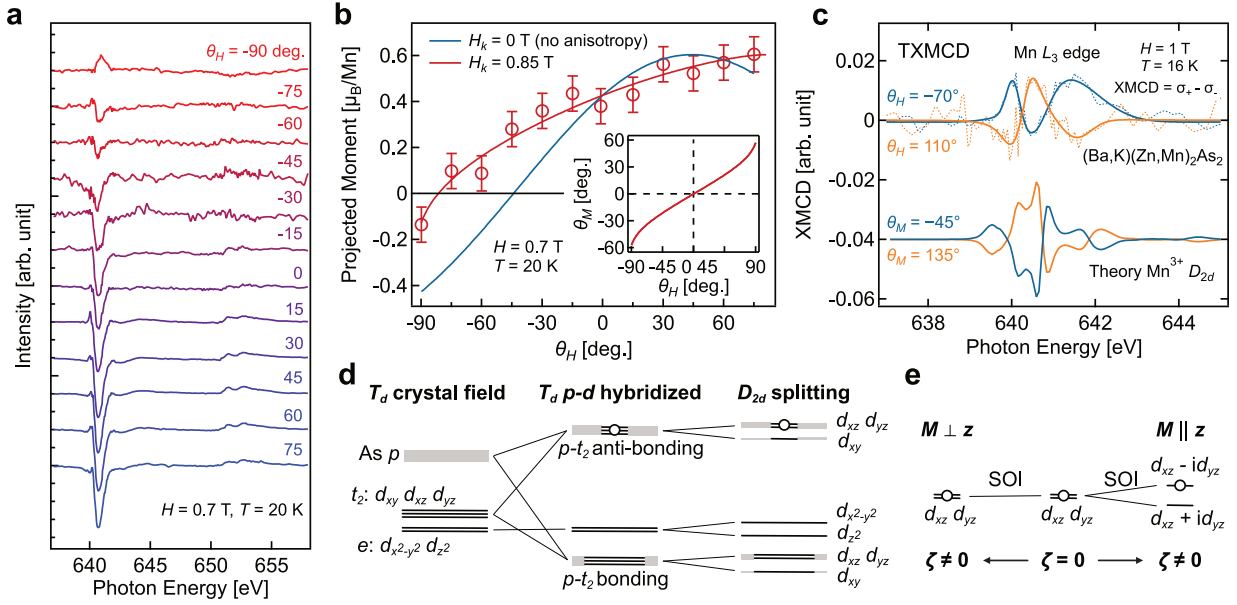}
\caption{{\bf a, b} Magnetic-field-angle dependence of XMCD spectra and Mn magnetic moment projected onto the incident light direction. In panel {\bf b}, the solid curves represent the results of the simulation using the Stoner-Wohlfarth model, and the inset shows the simulated magnetic-moment direction as a function of the magnetic-field angle. {\bf c} Transverse XMCD spectra measured with positive and negative transverse magnetic fields. Calculated spectra are also shown at the bottom. {\bf d} Schematic energy diagram of the Mn 3$d$ orbitals in (Ba,K)(Zn,Mn)$_{2}$As$_{2}$. {\bf e} Schematic energy diagram showing how the perpendicular magnetic anisotropy emerges as a result of the degeneracy lifting caused by spin-orbit interaction $\zeta$.}
\label{AD-XMCD}
\end{center}
\end{figure*}


Figures \ref{AD-XMCD}{\bf a} and \ref{AD-XMCD}{\bf b} show the magnetic-field-angle dependence of XMCD spectra and the total magnetic moments of Mn projected onto the incident light direction, respectively. Here, the magnetic moments are deduced using the XMCD sum rules \cite{Thole:1992aa,Carra:1993aa}. 
If there was no magnetic anisotropy and magnetic moments always pointed to the magnetic field direction, the data would follow a sine curve as shown by the blue curve in Fig. \ref{AD-XMCD}{\bf b}. However, the data clearly deviate from the sine curve, indicating a considerable magnetic anisotropy in this system. 
Here, we reproduce the data using the Stoner-Wohlfarth model. In this model, the total energy of the system is expressed as [3]
\begin{eqnarray}
E  = - \mu_{0}M_{\rm sat}&H&\cos{(\theta_{M} - \theta_{H})} \nonumber\\
& + & \frac{\mu_{0}}{2}M_{\rm sat}^{2}\cos^2{\theta_{M}} -  K_{U}\cos^{2}{\theta_{M}},
\end{eqnarray}
where $\mu_{0}$ denotes the permeability of vacuum, $M_{\rm sat}$ the saturation magnetization, $H$ the magnitude of the magnetic field and $K_{U}$ the uniaxial magneto-crystalline anisotropy energy per unit volume. As shown in Fig. \ref{Crystal}{\bf c}, $\theta_{M}$ and $\theta_{H}$ represent the angles of the magnetic moment and the magnetic field relative to the sample normal ($c$-axis direction). The first term represents the Zeeman energy, the second term the shape anisotropy energy, and the third term the uniaxial anisotropy energy. From this formula, one can calculate $\theta_{M}$ [or the projected moment $M_{\rm sat}$ $\cos(45^{\circ}-\theta_{M})$] for given $\theta_{H}$, $H$, $M_{\rm sat}$ and $K_{U}$ by minimizing the total energy $E$. In this way, we have fitted the data treating $K$ and $M_{\rm sat}$ as free parameters, and results are shown by a red curve in Fig. \ref{AD-XMCD}{\bf b}. The fit has yielded $K_{U} = (6.2 \pm 0.5) \times 10^4$ J/m$^3$ and the saturation magnetization per Mn atom $m_{\rm Mn}$ = 0.60 $\pm$ 0.03 $\mu_{\rm B}$. These values give the anisotropy field $2K_{U}/M_{\rm sat}$ of $0.85 \pm 0.07$ T. 
The positive value of $K_{U}$ means that the easy axis is along the $c$-axis, being consistent with the previous study \cite{Wang:2017aa}.
The obtained $m_{\rm Mn}$ of 0.60 $\mu_{\rm B}$ at 20 K is by far smaller than those of (Ga,Mn)As of 4.5 $\mu_{\rm B}$ [4]. This implies the existence of antiferromagnetically coupled Mn pairs or magnetically inactive Mn atoms, but this issue should be resolved in future studies.

Figure \ref{AD-XMCD}{\bf c} shows the TXMCD spectra, namely, XMCD spectra taken with the transverse geometry ($\theta_{H} = -70^{\circ}$, $110^{\circ}$ or $\theta_{M} \sim -45^{\circ}$, $135^{\circ}$) at which the spin component in the projected magnetic moment disappears and only the magnetic-dipole term is present. 
For the numerical correspondence between $\theta_{M}$ and $\theta_{H}$ in the TXMCD geometry, refer to the inset of Fig. \ref{AD-XMCD}{\bf b}.
The angles of the magnetic field were determined such that the XMCD intensity was minimized. In Fig. \ref{AD-XMCD}{\bf c}, the dashed curves represent the actual data and solid curves are guides to the eye, which are obtained by curve fitting with three Voigt functions.
The spectral line shape is very different from that of LXMCD, indicating that the signals observed here is not due to the residual spin component.
Moreover, the sign of the TXMCD spectra is reversed by rotating the magnetic field by 180$^{\circ}$. 
If these weak TXMCD spectra were just differential XAS spectra resulting from the slight photon-energy difference between left- and right-circularly polarized x rays, the two TXMCD spectra taken with the different magnetic-field directions should coincide.
Therefore, the sign reversal observed here is strong evidence to prove that the TXMCD signals are not artifacts but real.

The calculated TXMCD spectra are also shown at the bottom of Fig. \ref{AD-XMCD}{\bf c}. Although there are some discrepancies in the line shapes, the overall features well capture the experimental observation. According to the calculation, 0.2 eV $D_{2d}$ splitting resulted in almost fully occupied ($\sim$98\% filled) $d_{xy}$, $d_{z^{2}}$, $d_{x^2-y^2}$ orbitals and slightly less occupied ($\sim$89\% filled) $d_{xz}$, $d_{yz}$ orbitals. Therefore, holes are predominantly doped into the $d_{xz}$ and $d_{yz}$ orbitals, or $p$-$d_{xz}$ and $p$-$d_{yz}$ hybridized orbitals. 

This situation is schematically depicted in Fig. \ref{AD-XMCD}{\bf d}. Under the tetrahedral crystal field, the five Mn 3$d$ orbitals are split into doubly degenerate $e$ ($d_{x^2 - y^2}$, $d_{z^2}$) orbitals and triply degenerate $t_{2}$ ($d_{xy}, d_{xz}, d_{yz}$) orbitals as shown on the left-hand side of Fig. \ref{AD-XMCD}{\bf d}. The $t_{2}$ orbitals strongly hybridize with As 4$p$ orbitals and form bonding and anti-bonding $p$-$t_{2}$ orbitals, while the $e$ orbitals remain intact as shown in the middle column of Fig. \ref{AD-XMCD}{\bf d}. The bonding and anti-bonding $p$-$t_{2}$ hybridized orbitals predominantly consist of $t_{2}$ and $p$ orbitals, respectively. This orbital configuration is realized in cubic (Ga,Mn)As, and the holes residing in the anti-bonding $p$-$t_{2}$ hybridized states are the source of ferromagnetic exchange interaction. The elongation or compression of the MnAs$_{4}$ tetrahedra along the $c$-axis splits each of the $t_{2}$ and $e$ energy levels further into sub-levels: the $t_{2}$ level split into ($d_{xz}$, $d_{yz}$) and $d_{xy}$ levels, and the $e$ level split into $d_{x^2 - y^2}$ and $d_{z^2}$ levels. In the present system, the $d_{xz}$ and $d_{yz}$ levels lie higher in energy than the $d_{xy}$ level, and the $d_{x^2 - y^2}$ level higher than the $d_{z^2}$ level \cite{Gu:2016aa}, as shown on the right-hand side of Fig. \ref{AD-XMCD}{\bf d}. The doped holes thus reside in the $p$-$d_{xz}$ and $p$-$d_{yz}$ hybridized anti-bonding orbitals.

Magneto-crystalline anisotropy arises as a consequence of the energy gain of electrons occupying crystal-field-split orbitals caused by spin-orbit coupling when spins are aligned along a certain crystallographic direction, and only the orbitals near the Fermi level are relevant.
In the present system, the $p$-$d_{xz}$ and $p$-$d_{yz}$ hybridized anti-bonding orbitals near the Fermi level with holes in them should be responsible for the magnetic anisotropy. Figure \ref{AD-XMCD}{\bf e} shows how the partially occupied $d_{xz}$ and $d_{yz}$ orbitals can give rise to perpendicular ($z$-axis) magnetic anisotropy. When spins are aligned along the $z$-axis by a magnetic field, the degeneracy of the $d_{xz}$ and $d_{yz}$ orbitals will be lifted due to spin-orbit interaction to form $d_{xz} \pm id_{yz}$ $(L_{z}=\pm 1)$ orbitals, resulting in an energy gain. On the other hand, when spins are aligned in the $x$-$y$ plane, the $d_{xz}$ and $d_{yz}$ orbitals remain degenerate because any linear combination of these orbitals cannot form the eigenstate of $L_x$ or $L_y$, and thus there is no energy gain. 
This anisotropy of orbital magnetic moment may explain the difference in out-of-plane and in-plane saturation magnetizations observed in the previous study \cite{Zhao:2017aa}.
This situation is similar to the cases of Fe/MgO \cite{Yang:2011aa, Okabayashi:2014aa} and Co/Pt interfaces \cite{Nakajima:1998aa}, where the origin of their large PMA was attributed to the degeneracy lifting of the $d_{xz}$ and $d_{yz}$ orbitals near the Fermi level. 

The present results also imply that it is possible to control the magnetic anisotropy by changing the number of carriers to change the electron occupation of each $d$ orbital or even by the isoelectric substitutions that changes the magnitude of the $D_{2d}$ splitting. 
These degrees of freedom would enable one to independently control the Curie temperature, carrier concentration, and magnetic anisotropy, which would be useful for future spintronics applications. Such potential functionalities should be explored in future studies.

In summary, we have performed angle-dependent XMCD study to reveal the origin of perpendicular magnetic anisotropy of (Ba,K)(Zn,Mn)$_{2}$As$_{2}$. 
Using the Stoner-Wohlfarth model fitting, the magnetic anisotropy energy was estimated to be $K_{U} = (6.2 \pm 0.5) \times 10^{4}$ J/m$^3$  and the saturation magnetization per Mn atom was estimated to be $m_{\rm Mn}$ = 0.60 $\pm$ 0.03 $\mu_{\rm B}$. We have observed transverse XMCD spectra, which have been well reproduced by cluster-model calculation with $D_{2d}$ splitting where holes reside in the $d_{xz}$ and $d_{yz}$ orbitals. 
We conclude that the magnetic anisotropy originate from the degeneracy lifting of those orbitals due to spin-orbit coupling and resulting energy gain when spins are aligned along the $z$-direction.

\begin{acknowledgments}
We would like to thank Kenta Amemiya  and Masako Sakamaki for valuable technical support at Photon Factory, KEK.
This work was supported by Grants-in-Aid for Scientific Research from the JSPS (Grants No. 15H02109, No. 15K17696, and 19K03741).
The experiment was done under the approval of the Program Advisory Committee (Proposal No. 2016S2-005, 2016G066).
The works at IOPCAS are supported by NSF \& MOST of China through research projects.
The work at Columbia was supported by the US NSF DMR1610633.
S.S. acknowledges financial support from Advanced Leading Graduate Course for Photon Science (ALPS) and the JSPS Research Fellowship for Young Scientists. 
A.F. acknowledges support as an adjunct member of the Center for Spintronics Research Network (CSRN), the University of Tokyo, under the Spintronics Research Network of Japan (Spin-RNJ).
\end{acknowledgments}



\newpage
\section{Supplementary Information}



\subsection{Sample growth}
Ba$_{0.904}$K$_{0.096}$(Zn$_{0.805}$Mn$_{0.195}$)$_{2}$As$_{2}$ single crystals with $T_{\rm C}$ = 60 K were grown by the flux technique. (Zn,Mn)As precursors were first prepared by heating the mixture of high-purity Zn, Mn, and As at 750 ℃ for 35 hours. Ba and K were then incorporated into a quartz tube with the precursors.
The quartz tube was heated at 1200 ℃ for 48 hours and cooled down to room temperature at the rate of 3 ℃/h. See ref. \cite{Zhao:2017aa} for further details.

\color{black}
\subsection{Cluster-model calculations}

\subsubsection*{Computational details}
We used the code Xtls version 8.5 for the CI cluster-model calculation \cite{Tanaka:1994aa}. 
In the calculation, we have employed the same geometry as the experiment. We have assumed a tetrahedral [MnAs$_{4}$]$^{-9}$ cluster (Mn$^{3+}$ cluster), the ground state of which is represented by the superposition of $d^4$, $d^5\underline{L}$, and $d^6\underline{L}^2$ configurations, where $\underline{L}$ denotes a ligand hole. The Mn$^{3+}$ cluster rather than the Mn$^{2+}$ cluster consisting of $d^5$, $d^6\underline{L}$, and $d^7\underline{L}^2$ configurations was reported to better reproduce the resonant inelastic x-ray scattering spectra of (Ga,Mn)As \cite{Kobayashi:2016aa}. Note that the Mn$^{3+}$ cluster has the predominant $d^5\underline{L}$ configuration, and hence the difference from the Mn$^{2+}$ cluster whose ground state has the predominant $d^5$ configuration is rather subtle. We have used basically the same parameters as those used for (Ga,Mn)As \cite{Kobayashi:2016aa} because the Mn atoms in (Ba,K)(Zn,Mn)$_{2}$As$_{2}$ are coordinated by the distorted tetrahedra As$_{4}$, similar to those in (Ga,Mn)As coordinated by the tetrahedra As$_{4}$. 
The ligand-to-3$d$ charge-transfer energy $\Delta$, defined as the energy difference between the $d^{5}\underline{L}$ and $d^{6}\underline{L}^{2}$ states, was set to $\Delta = 1.5$ eV, the $d$-$d$ Coulomb interaction energy $U_{dd} = 3.5$ eV, the Slater-Koster parameter $pd\sigma = - 0.9$ eV, and the tetrahedral crystal field $-10Dq_{\rm crys} = 0$ eV. Although $-10Dq_{\rm crys}$ was set to be 0 eV, the finite $p$-$d$ hybridization causes effective $T_{d}$ crystal-field splitting of $-10Dq_{\rm hyb} \sim -pd\sigma$ \cite{Kobayashi:2016aa}. 
The Slater integrals for the 3$d$-3$d$ and 3$d$-2$p$ multipole interactions are reduced to 80\% of the atomic Hartree-Fock values, and the 2$p$ spin-orbit interaction is scaled to 103\% \cite{Freeman:2006aa}. In order to model the elongated tetrahedra, or the quasi two-dimensional crystal structure, we have introduced an additional small $D_{2d}$ splitting that makes the $d_{xz}$ and $d_{yz}$ orbitals lie higher in energy by 0.2 eV than the $d_{xy}$ orbital, and the $d_{x^2-y^2}$ orbital higher in energy by 0.2 eV than $d_{z^2}$. The parameters for $D_{2d}$ splitting were chosen based on the DFT calculation \cite{Gu:2016aa} which showed a $\sim$0.2 eV splitting for the relevant orbitals. The calculated ground state consists of 9\% $d^{4}$, 76\% $d^{6}{\underline L}^2$, and 15\% $d^{6}{\underline L}^2$ configurations.
Note that the $D_{2d}$ splitting between the $d_{z^2}$ and $d_{x^2-y^2}$ orbitals adopted in the present calculations is opposite to what is expected from $D_{2d}$ {\it crystal field} in an elongated tetrahedron. This counterintuitive orbital configuration is probably resulted from the network of edge-shared tetrahedra, where there would be additional crystal field from neighboring tetrahedra.

\subsubsection*{Calculations with various $D_{2d}$ parameters}

\begin{figure}
\begin{center}
\includegraphics[width=17cm]{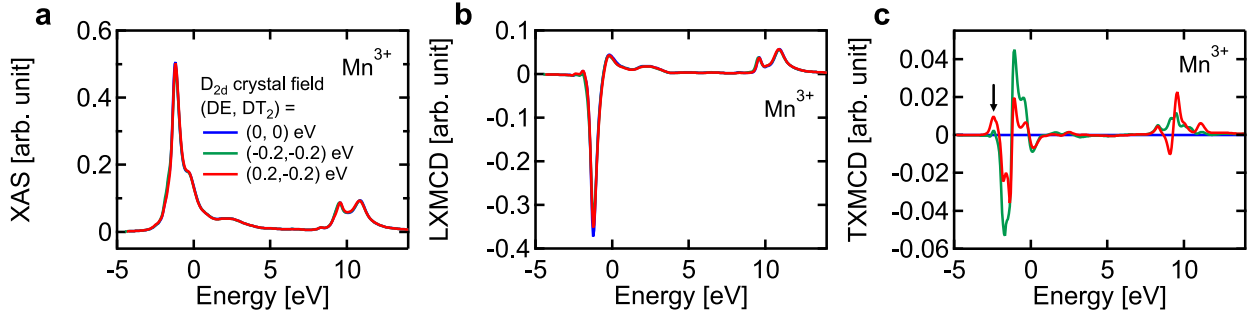}
\caption{Cluster model calculations. {\bf a} XAS, {\bf b} LXMCD, and {\bf c} TXMCD spectra for Mn$^{3+}$ [MnAs$_{4}$]$^{-9}$ cluster with or without $D_{2d}$ splitting.}
\label{Cluster}
\end{center}
\end{figure}

Figure \ref{Cluster} shows the calculated XAS and XMCD spectra with or without $D_{2d}$ splitting in longitudinal and transverse geometry, referred to as LXMCD and TXMCD, respectively. In longitudinal (transverse) geometry, the magnetic moment is aligned parallel (perpendicular) to the incident x-ray direction.
The $D_{2d}$ splitting makes $t_{2g}$ orbitals split into $d_{xz,yz}$ and $d_{xy}$ orbitals and $e_{g}$ orbitals split into $d_{z^2}$ and $d_{x^2-y^2}$ orbitals. The energy difference between $d_{xy}$ and ($d_{xz}$, $d_{yz}$) orbitals ($E_{xy} -E_{xz,yz}$) is denoted as $DT_{2}$, and the energy difference between $d_{z^2}$ and $d_{x^2-y^2}$ orbitals ($E_{x^2-y^2} -E_{z^2}$) is denoted as $DE$. 
While the XAS and LXMCD spectra are rather insensitive to the $D_{2d}$ splitting, the TXMCD spectra change their line shape: only with the positive $DE = 0.2$ eV, a positive peak appears below the XAS peak position as indicated by black arrows in fig. \ref{Cluster}{\bf c}. 
Because such a pre-edge peak was also observed in the experiment as shown in fig. 3{\bf c} in the main text, it can be said that the $d_{x^2-y^2}$ orbital lies higher in energy than the $d_{z^2}$ orbitals. As mentioned above, the positive sign of $DE$ is opposite to what is expected from $D_{2d}$ {\it crystal field} in a lone elongated tetrahedron but agrees with the previous theoretical calculations \cite{Gu:2016aa}.

\bibliography{Bibtex_BKZMA}

\begin{thebibliography}{40}%
\makeatletter
\providecommand \@ifxundefined [1]{%
 \@ifx{#1\undefined}
}%
\providecommand \@ifnum [1]{%
 \ifnum #1\expandafter \@firstoftwo
 \else \expandafter \@secondoftwo
 \fi
}%
\providecommand \@ifx [1]{%
 \ifx #1\expandafter \@firstoftwo
 \else \expandafter \@secondoftwo
 \fi
}%
\providecommand \natexlab [1]{#1}%
\providecommand \enquote  [1]{``#1''}%
\providecommand \bibnamefont  [1]{#1}%
\providecommand \bibfnamefont [1]{#1}%
\providecommand \citenamefont [1]{#1}%
\providecommand \href@noop [0]{\@secondoftwo}%
\providecommand \href [0]{\begingroup \@sanitize@url \@href}%
\providecommand \@href[1]{\@@startlink{#1}\@@href}%
\providecommand \@@href[1]{\endgroup#1\@@endlink}%
\providecommand \@sanitize@url [0]{\catcode `\\12\catcode `\$12\catcode
  `\&12\catcode `\#12\catcode `\^12\catcode `\_12\catcode `\%12\relax}%
\providecommand \@@startlink[1]{}%
\providecommand \@@endlink[0]{}%
\providecommand \url  [0]{\begingroup\@sanitize@url \@url }%
\providecommand \@url [1]{\endgroup\@href {#1}{\urlprefix }}%
\providecommand \urlprefix  [0]{URL }%
\providecommand \Eprint [0]{\href }%
\providecommand \doibase [0]{http://dx.doi.org/}%
\providecommand \selectlanguage [0]{\@gobble}%
\providecommand \bibinfo  [0]{\@secondoftwo}%
\providecommand \bibfield  [0]{\@secondoftwo}%
\providecommand \translation [1]{[#1]}%
\providecommand \BibitemOpen [0]{}%
\providecommand \bibitemStop [0]{}%
\providecommand \bibitemNoStop [0]{.\EOS\space}%
\providecommand \EOS [0]{\spacefactor3000\relax}%
\providecommand \BibitemShut  [1]{\csname bibitem#1\endcsname}%
\let\auto@bib@innerbib\@empty
\bibitem [{\citenamefont {Ohno}(1998)}]{Ohno:1998aa}%
  \BibitemOpen
  \bibfield  {author} {\bibinfo {author} {\bibfnamefont {H.}~\bibnamefont
  {Ohno}},\ }\href {\doibase 10.1126/science.281.5379.951} {\bibfield
  {journal} {\bibinfo  {journal} {Science}\ }\textbf {\bibinfo {volume}
  {281}},\ \bibinfo {pages} {951} (\bibinfo {year} {1998})}\BibitemShut
  {NoStop}%
\bibitem [{\citenamefont {Munekata}\ \emph {et~al.}(1989)\citenamefont
  {Munekata}, \citenamefont {Ohno}, \citenamefont {von Molnar}, \citenamefont
  {Segm\"uller}, \citenamefont {Chang},\ and\ \citenamefont
  {Esaki}}]{Munekata:1989aa}%
  \BibitemOpen
  \bibfield  {author} {\bibinfo {author} {\bibfnamefont {H.}~\bibnamefont
  {Munekata}}, \bibinfo {author} {\bibfnamefont {H.}~\bibnamefont {Ohno}},
  \bibinfo {author} {\bibfnamefont {S.}~\bibnamefont {von Molnar}}, \bibinfo
  {author} {\bibfnamefont {A.}~\bibnamefont {Segm\"uller}}, \bibinfo {author}
  {\bibfnamefont {L.~L.}\ \bibnamefont {Chang}}, \ and\ \bibinfo {author}
  {\bibfnamefont {L.}~\bibnamefont {Esaki}},\ }\href {\doibase
  10.1103/PhysRevLett.63.1849} {\bibfield  {journal} {\bibinfo  {journal}
  {Phys. Rev. Lett.}\ }\textbf {\bibinfo {volume} {63}},\ \bibinfo {pages}
  {1849} (\bibinfo {year} {1989})}\BibitemShut {NoStop}%
\bibitem [{\citenamefont {Dietl}\ and\ \citenamefont
  {Ohno}(2014)}]{Dietl:2014aa}%
  \BibitemOpen
  \bibfield  {author} {\bibinfo {author} {\bibfnamefont {T.}~\bibnamefont
  {Dietl}}\ and\ \bibinfo {author} {\bibfnamefont {H.}~\bibnamefont {Ohno}},\
  }\href {\doibase 10.1103/RevModPhys.86.187} {\bibfield  {journal} {\bibinfo
  {journal} {Rev. Mod. Phys.}\ }\textbf {\bibinfo {volume} {86}},\ \bibinfo
  {pages} {187} (\bibinfo {year} {2014})}\BibitemShut {NoStop}%
\bibitem [{\citenamefont {Jungwirth}\ \emph {et~al.}(2014)\citenamefont
  {Jungwirth}, \citenamefont {Wunderlich}, \citenamefont {Nov{\'a}k},
  \citenamefont {Olejn{\'\i}k}, \citenamefont {Gallagher}, \citenamefont
  {Campion}, \citenamefont {Edmonds}, \citenamefont {Rushforth}, \citenamefont
  {Ferguson},\ and\ \citenamefont {N{\v e}mec}}]{Jungwirth:2014aa}%
  \BibitemOpen
  \bibfield  {author} {\bibinfo {author} {\bibfnamefont {T.}~\bibnamefont
  {Jungwirth}}, \bibinfo {author} {\bibfnamefont {J.}~\bibnamefont
  {Wunderlich}}, \bibinfo {author} {\bibfnamefont {V.}~\bibnamefont
  {Nov{\'a}k}}, \bibinfo {author} {\bibfnamefont {K.}~\bibnamefont
  {Olejn{\'\i}k}}, \bibinfo {author} {\bibfnamefont {B.~L.}\ \bibnamefont
  {Gallagher}}, \bibinfo {author} {\bibfnamefont {R.~P.}\ \bibnamefont
  {Campion}}, \bibinfo {author} {\bibfnamefont {K.~W.}\ \bibnamefont
  {Edmonds}}, \bibinfo {author} {\bibfnamefont {A.~W.}\ \bibnamefont
  {Rushforth}}, \bibinfo {author} {\bibfnamefont {A.~J.}\ \bibnamefont
  {Ferguson}}, \ and\ \bibinfo {author} {\bibfnamefont {P.}~\bibnamefont {N{\v
  e}mec}},\ }\href {\doibase 10.1103/RevModPhys.86.855} {\bibfield  {journal}
  {\bibinfo  {journal} {Rev. Mod. Phys.}\ }\textbf {\bibinfo {volume} {86}},\
  \bibinfo {pages} {855} (\bibinfo {year} {2014})}\BibitemShut {NoStop}%
\bibitem [{\citenamefont {Zhao}\ \emph {et~al.}(2013)\citenamefont {Zhao},
  \citenamefont {Deng}, \citenamefont {Wang}, \citenamefont {Han},
  \citenamefont {Zhu}, \citenamefont {Li}, \citenamefont {Liu}, \citenamefont
  {Yu}, \citenamefont {Goko}, \citenamefont {Frandsen}, \citenamefont {Liu},
  \citenamefont {Ning}, \citenamefont {Uemura}, \citenamefont {Dabkowska},
  \citenamefont {Luke}, \citenamefont {Luetkens}, \citenamefont {Morenzoni},
  \citenamefont {Dunsiger}, \citenamefont {Senyshyn}, \citenamefont
  {B{\"o}ni},\ and\ \citenamefont {Jin}}]{Zhao:2013aa}%
  \BibitemOpen
  \bibfield  {author} {\bibinfo {author} {\bibfnamefont {K.}~\bibnamefont
  {Zhao}}, \bibinfo {author} {\bibfnamefont {Z.}~\bibnamefont {Deng}}, \bibinfo
  {author} {\bibfnamefont {X.~C.}\ \bibnamefont {Wang}}, \bibinfo {author}
  {\bibfnamefont {W.}~\bibnamefont {Han}}, \bibinfo {author} {\bibfnamefont
  {J.~L.}\ \bibnamefont {Zhu}}, \bibinfo {author} {\bibfnamefont
  {X.}~\bibnamefont {Li}}, \bibinfo {author} {\bibfnamefont {Q.~Q.}\
  \bibnamefont {Liu}}, \bibinfo {author} {\bibfnamefont {R.~C.}\ \bibnamefont
  {Yu}}, \bibinfo {author} {\bibfnamefont {T.}~\bibnamefont {Goko}}, \bibinfo
  {author} {\bibfnamefont {B.}~\bibnamefont {Frandsen}}, \bibinfo {author}
  {\bibfnamefont {L.}~\bibnamefont {Liu}}, \bibinfo {author} {\bibfnamefont
  {F.}~\bibnamefont {Ning}}, \bibinfo {author} {\bibfnamefont {Y.~J.}\
  \bibnamefont {Uemura}}, \bibinfo {author} {\bibfnamefont {H.}~\bibnamefont
  {Dabkowska}}, \bibinfo {author} {\bibfnamefont {G.}~\bibnamefont {Luke}},
  \bibinfo {author} {\bibfnamefont {H.}~\bibnamefont {Luetkens}}, \bibinfo
  {author} {\bibfnamefont {E.}~\bibnamefont {Morenzoni}}, \bibinfo {author}
  {\bibfnamefont {S.~R.}\ \bibnamefont {Dunsiger}}, \bibinfo {author}
  {\bibfnamefont {A.}~\bibnamefont {Senyshyn}}, \bibinfo {author}
  {\bibfnamefont {P.}~\bibnamefont {B{\"o}ni}}, \ and\ \bibinfo {author}
  {\bibfnamefont {C.~Q.}\ \bibnamefont {Jin}},\ }\href@noop {} {\bibfield
  {journal} {\bibinfo  {journal} {Nat. Commun.}\ }\textbf {\bibinfo {volume}
  {4}},\ \bibinfo {pages} {1442} (\bibinfo {year} {2013})}\BibitemShut
  {NoStop}%
\bibitem [{\citenamefont {Zhao}\ \emph {et~al.}(2017)\citenamefont {Zhao},
  \citenamefont {Lin}, \citenamefont {Deng}, \citenamefont {Gu}, \citenamefont
  {Yu}, \citenamefont {Wang}, \citenamefont {Gong}, \citenamefont {Uemera},
  \citenamefont {Li},\ and\ \citenamefont {Jin}}]{Zhao:2017aa}%
  \BibitemOpen
  \bibfield  {author} {\bibinfo {author} {\bibfnamefont {G.~Q.}\ \bibnamefont
  {Zhao}}, \bibinfo {author} {\bibfnamefont {C.~J.}\ \bibnamefont {Lin}},
  \bibinfo {author} {\bibfnamefont {Z.}~\bibnamefont {Deng}}, \bibinfo {author}
  {\bibfnamefont {G.~X.}\ \bibnamefont {Gu}}, \bibinfo {author} {\bibfnamefont
  {S.}~\bibnamefont {Yu}}, \bibinfo {author} {\bibfnamefont {X.~C.}\
  \bibnamefont {Wang}}, \bibinfo {author} {\bibfnamefont {Z.~Z.}\ \bibnamefont
  {Gong}}, \bibinfo {author} {\bibfnamefont {Y.~J.}\ \bibnamefont {Uemera}},
  \bibinfo {author} {\bibfnamefont {Y.~Q.}\ \bibnamefont {Li}}, \ and\ \bibinfo
  {author} {\bibfnamefont {C.~Q.}\ \bibnamefont {Jin}},\ }\href {\doibase
  10.1038/s41598-017-08394-z} {\bibfield  {journal} {\bibinfo  {journal}
  {Scientific Reports}\ }\textbf {\bibinfo {volume} {7}},\ \bibinfo {pages}
  {14473} (\bibinfo {year} {2017})}\BibitemShut {NoStop}%
\bibitem [{\citenamefont {Xiao}\ \emph {et~al.}(2014)\citenamefont {Xiao},
  \citenamefont {Ran}, \citenamefont {Hiramatsu}, \citenamefont {Matsuishi},
  \citenamefont {Hosono},\ and\ \citenamefont {Kamiya}}]{Xiao:2014ab}%
  \BibitemOpen
  \bibfield  {author} {\bibinfo {author} {\bibfnamefont {Z.}~\bibnamefont
  {Xiao}}, \bibinfo {author} {\bibfnamefont {F.-Y.}\ \bibnamefont {Ran}},
  \bibinfo {author} {\bibfnamefont {H.}~\bibnamefont {Hiramatsu}}, \bibinfo
  {author} {\bibfnamefont {S.}~\bibnamefont {Matsuishi}}, \bibinfo {author}
  {\bibfnamefont {H.}~\bibnamefont {Hosono}}, \ and\ \bibinfo {author}
  {\bibfnamefont {T.}~\bibnamefont {Kamiya}},\ }\href {\doibase
  http://dx.doi.org/10.1016/j.tsf.2013.10.135} {\bibfield  {journal} {\bibinfo
  {journal} {Thin Solid Films}\ }\textbf {\bibinfo {volume} {559}},\ \bibinfo
  {pages} {100 } (\bibinfo {year} {2014})}\BibitemShut {NoStop}%
\bibitem [{\citenamefont {Zhao}\ \emph {et~al.}(2014)\citenamefont {Zhao},
  \citenamefont {Chen}, \citenamefont {Zhao}, \citenamefont {Yuan},
  \citenamefont {Liu}, \citenamefont {Deng}, \citenamefont {Zhu},\ and\
  \citenamefont {Jin}}]{Zhao:2014aa}%
  \BibitemOpen
  \bibfield  {author} {\bibinfo {author} {\bibfnamefont {K.}~\bibnamefont
  {Zhao}}, \bibinfo {author} {\bibfnamefont {B.}~\bibnamefont {Chen}}, \bibinfo
  {author} {\bibfnamefont {G.}~\bibnamefont {Zhao}}, \bibinfo {author}
  {\bibfnamefont {Z.}~\bibnamefont {Yuan}}, \bibinfo {author} {\bibfnamefont
  {Q.}~\bibnamefont {Liu}}, \bibinfo {author} {\bibfnamefont {Z.}~\bibnamefont
  {Deng}}, \bibinfo {author} {\bibfnamefont {J.}~\bibnamefont {Zhu}}, \ and\
  \bibinfo {author} {\bibfnamefont {C.}~\bibnamefont {Jin}},\ }\href {\doibase
  10.1007/s11434-014-0398-z} {\bibfield  {journal} {\bibinfo  {journal} {Chin.
  Sci. Bull.}\ }\textbf {\bibinfo {volume} {59}},\ \bibinfo {pages} {2524}
  (\bibinfo {year} {2014})}\BibitemShut {NoStop}%
\bibitem [{\citenamefont {Chen}\ \emph {et~al.}(2011)\citenamefont {Chen},
  \citenamefont {Yang}, \citenamefont {Yang}, \citenamefont {Zhao},
  \citenamefont {Misuraca}, \citenamefont {Xiong},\ and\ \citenamefont {von
  Moln{\'a}r}}]{Chen:2011aa}%
  \BibitemOpen
  \bibfield  {author} {\bibinfo {author} {\bibfnamefont {L.}~\bibnamefont
  {Chen}}, \bibinfo {author} {\bibfnamefont {X.}~\bibnamefont {Yang}}, \bibinfo
  {author} {\bibfnamefont {F.}~\bibnamefont {Yang}}, \bibinfo {author}
  {\bibfnamefont {J.}~\bibnamefont {Zhao}}, \bibinfo {author} {\bibfnamefont
  {J.}~\bibnamefont {Misuraca}}, \bibinfo {author} {\bibfnamefont
  {P.}~\bibnamefont {Xiong}}, \ and\ \bibinfo {author} {\bibfnamefont
  {S.}~\bibnamefont {von Moln{\'a}r}},\ }\href {\doibase 10.1021/nl201187m}
  {\bibfield  {journal} {\bibinfo  {journal} {Nano Lett.}\ }\textbf {\bibinfo
  {volume} {11}},\ \bibinfo {pages} {2584} (\bibinfo {year}
  {2011})}\BibitemShut {NoStop}%
\bibitem [{\citenamefont {Sun}\ \emph {et~al.}(2017)\citenamefont {Sun},
  \citenamefont {Zhao}, \citenamefont {Escanhoela}, \citenamefont {Chen},
  \citenamefont {Kou}, \citenamefont {Wang}, \citenamefont {Xiao},
  \citenamefont {Chow}, \citenamefont {Mao}, \citenamefont {Haskel},
  \citenamefont {Yang},\ and\ \citenamefont {Jin}}]{Sun:2017aa}%
  \BibitemOpen
  \bibfield  {author} {\bibinfo {author} {\bibfnamefont {F.}~\bibnamefont
  {Sun}}, \bibinfo {author} {\bibfnamefont {G.~Q.}\ \bibnamefont {Zhao}},
  \bibinfo {author} {\bibfnamefont {C.~A.}\ \bibnamefont {Escanhoela}},
  \bibinfo {author} {\bibfnamefont {B.~J.}\ \bibnamefont {Chen}}, \bibinfo
  {author} {\bibfnamefont {R.~H.}\ \bibnamefont {Kou}}, \bibinfo {author}
  {\bibfnamefont {Y.~G.}\ \bibnamefont {Wang}}, \bibinfo {author}
  {\bibfnamefont {Y.~M.}\ \bibnamefont {Xiao}}, \bibinfo {author}
  {\bibfnamefont {P.}~\bibnamefont {Chow}}, \bibinfo {author} {\bibfnamefont
  {H.~K.}\ \bibnamefont {Mao}}, \bibinfo {author} {\bibfnamefont
  {D.}~\bibnamefont {Haskel}}, \bibinfo {author} {\bibfnamefont {W.~G.}\
  \bibnamefont {Yang}}, \ and\ \bibinfo {author} {\bibfnamefont {C.~Q.}\
  \bibnamefont {Jin}},\ }\href {\doibase 10.1103/PhysRevB.95.094412} {\bibfield
   {journal} {\bibinfo  {journal} {Phys. Rev. B}\ }\textbf {\bibinfo {volume}
  {95}},\ \bibinfo {pages} {094412} (\bibinfo {year} {2017})}\BibitemShut
  {NoStop}%
\bibitem [{\citenamefont {Zhao}\ \emph {et~al.}(2018)\citenamefont {Zhao},
  \citenamefont {Li}, \citenamefont {Sun}, \citenamefont {Yuan}, \citenamefont
  {Chen}, \citenamefont {Yu}, \citenamefont {Peng}, \citenamefont {Deng},
  \citenamefont {Wang},\ and\ \citenamefont {Jin}}]{Zhao:2018aa}%
  \BibitemOpen
  \bibfield  {author} {\bibinfo {author} {\bibfnamefont {G.~Q.}\ \bibnamefont
  {Zhao}}, \bibinfo {author} {\bibfnamefont {Z.}~\bibnamefont {Li}}, \bibinfo
  {author} {\bibfnamefont {F.}~\bibnamefont {Sun}}, \bibinfo {author}
  {\bibfnamefont {Z.}~\bibnamefont {Yuan}}, \bibinfo {author} {\bibfnamefont
  {B.~J.}\ \bibnamefont {Chen}}, \bibinfo {author} {\bibfnamefont
  {S.}~\bibnamefont {Yu}}, \bibinfo {author} {\bibfnamefont {Y.}~\bibnamefont
  {Peng}}, \bibinfo {author} {\bibfnamefont {Z.}~\bibnamefont {Deng}}, \bibinfo
  {author} {\bibfnamefont {X.~C.}\ \bibnamefont {Wang}}, \ and\ \bibinfo
  {author} {\bibfnamefont {C.~Q.}\ \bibnamefont {Jin}},\ }\href {\doibase
  10.1088/1361-648x/aac367} {\bibfield  {journal} {\bibinfo  {journal} {Journal
  of Physics: Condensed Matter}\ }\textbf {\bibinfo {volume} {30}},\ \bibinfo
  {pages} {254001} (\bibinfo {year} {2018})}\BibitemShut {NoStop}%
\bibitem [{\citenamefont {Suzuki}\ \emph
  {et~al.}(2015{\natexlab{a}})\citenamefont {Suzuki}, \citenamefont {Zhao},
  \citenamefont {Zhao}, \citenamefont {Chen}, \citenamefont {Horio},
  \citenamefont {Koshiishi}, \citenamefont {Xu}, \citenamefont {Kobayashi},
  \citenamefont {Minohara}, \citenamefont {Sakai}, \citenamefont {Horiba},
  \citenamefont {Kumigashira}, \citenamefont {Gu}, \citenamefont {Maekawa},
  \citenamefont {Uemura}, \citenamefont {Jin},\ and\ \citenamefont
  {Fujimori}}]{Suzuki:2015aa}%
  \BibitemOpen
  \bibfield  {author} {\bibinfo {author} {\bibfnamefont {H.}~\bibnamefont
  {Suzuki}}, \bibinfo {author} {\bibfnamefont {G.~Q.}\ \bibnamefont {Zhao}},
  \bibinfo {author} {\bibfnamefont {K.}~\bibnamefont {Zhao}}, \bibinfo {author}
  {\bibfnamefont {B.~J.}\ \bibnamefont {Chen}}, \bibinfo {author}
  {\bibfnamefont {M.}~\bibnamefont {Horio}}, \bibinfo {author} {\bibfnamefont
  {K.}~\bibnamefont {Koshiishi}}, \bibinfo {author} {\bibfnamefont
  {J.}~\bibnamefont {Xu}}, \bibinfo {author} {\bibfnamefont {M.}~\bibnamefont
  {Kobayashi}}, \bibinfo {author} {\bibfnamefont {M.}~\bibnamefont {Minohara}},
  \bibinfo {author} {\bibfnamefont {E.}~\bibnamefont {Sakai}}, \bibinfo
  {author} {\bibfnamefont {K.}~\bibnamefont {Horiba}}, \bibinfo {author}
  {\bibfnamefont {H.}~\bibnamefont {Kumigashira}}, \bibinfo {author}
  {\bibfnamefont {B.}~\bibnamefont {Gu}}, \bibinfo {author} {\bibfnamefont
  {S.}~\bibnamefont {Maekawa}}, \bibinfo {author} {\bibfnamefont {Y.~J.}\
  \bibnamefont {Uemura}}, \bibinfo {author} {\bibfnamefont {C.~Q.}\
  \bibnamefont {Jin}}, \ and\ \bibinfo {author} {\bibfnamefont
  {A.}~\bibnamefont {Fujimori}},\ }\href {\doibase 10.1103/PhysRevB.92.235120}
  {\bibfield  {journal} {\bibinfo  {journal} {Phys. Rev. B}\ }\textbf {\bibinfo
  {volume} {92}},\ \bibinfo {pages} {235120} (\bibinfo {year}
  {2015}{\natexlab{a}})}\BibitemShut {NoStop}%
\bibitem [{\citenamefont {Suzuki}\ \emph
  {et~al.}(2015{\natexlab{b}})\citenamefont {Suzuki}, \citenamefont {Zhao},
  \citenamefont {Shibata}, \citenamefont {Takahashi}, \citenamefont {Sakamoto},
  \citenamefont {Yoshimatsu}, \citenamefont {Chen}, \citenamefont
  {Kumigashira}, \citenamefont {Chang}, \citenamefont {Lin}, \citenamefont
  {Huang}, \citenamefont {Chen}, \citenamefont {Gu}, \citenamefont {Maekawa},
  \citenamefont {Uemura}, \citenamefont {Jin},\ and\ \citenamefont
  {Fujimori}}]{Suzuki:2015ab}%
  \BibitemOpen
  \bibfield  {author} {\bibinfo {author} {\bibfnamefont {H.}~\bibnamefont
  {Suzuki}}, \bibinfo {author} {\bibfnamefont {K.}~\bibnamefont {Zhao}},
  \bibinfo {author} {\bibfnamefont {G.}~\bibnamefont {Shibata}}, \bibinfo
  {author} {\bibfnamefont {Y.}~\bibnamefont {Takahashi}}, \bibinfo {author}
  {\bibfnamefont {S.}~\bibnamefont {Sakamoto}}, \bibinfo {author}
  {\bibfnamefont {K.}~\bibnamefont {Yoshimatsu}}, \bibinfo {author}
  {\bibfnamefont {B.~J.}\ \bibnamefont {Chen}}, \bibinfo {author}
  {\bibfnamefont {H.}~\bibnamefont {Kumigashira}}, \bibinfo {author}
  {\bibfnamefont {F.-H.}\ \bibnamefont {Chang}}, \bibinfo {author}
  {\bibfnamefont {H.-J.}\ \bibnamefont {Lin}}, \bibinfo {author} {\bibfnamefont
  {D.~J.}\ \bibnamefont {Huang}}, \bibinfo {author} {\bibfnamefont {C.~T.}\
  \bibnamefont {Chen}}, \bibinfo {author} {\bibfnamefont {B.}~\bibnamefont
  {Gu}}, \bibinfo {author} {\bibfnamefont {S.}~\bibnamefont {Maekawa}},
  \bibinfo {author} {\bibfnamefont {Y.~J.}\ \bibnamefont {Uemura}}, \bibinfo
  {author} {\bibfnamefont {C.~Q.}\ \bibnamefont {Jin}}, \ and\ \bibinfo
  {author} {\bibfnamefont {A.}~\bibnamefont {Fujimori}},\ }\href {\doibase
  10.1103/PhysRevB.91.140401} {\bibfield  {journal} {\bibinfo  {journal} {Phys.
  Rev. B}\ }\textbf {\bibinfo {volume} {91}},\ \bibinfo {pages} {140401}
  (\bibinfo {year} {2015}{\natexlab{b}})}\BibitemShut {NoStop}%
\bibitem [{\citenamefont {Sun}\ \emph {et~al.}(2016)\citenamefont {Sun},
  \citenamefont {Li}, \citenamefont {Chen}, \citenamefont {Jia}, \citenamefont
  {Zhang}, \citenamefont {Li}, \citenamefont {Zhao}, \citenamefont {Xing},
  \citenamefont {Fabbris}, \citenamefont {Wang}, \citenamefont {Deng},
  \citenamefont {Uemura}, \citenamefont {Mao}, \citenamefont {Haskel},
  \citenamefont {Yang},\ and\ \citenamefont {Jin}}]{Sun:2016aa}%
  \BibitemOpen
  \bibfield  {author} {\bibinfo {author} {\bibfnamefont {F.}~\bibnamefont
  {Sun}}, \bibinfo {author} {\bibfnamefont {N.~N.}\ \bibnamefont {Li}},
  \bibinfo {author} {\bibfnamefont {B.~J.}\ \bibnamefont {Chen}}, \bibinfo
  {author} {\bibfnamefont {Y.~T.}\ \bibnamefont {Jia}}, \bibinfo {author}
  {\bibfnamefont {L.~J.}\ \bibnamefont {Zhang}}, \bibinfo {author}
  {\bibfnamefont {W.~M.}\ \bibnamefont {Li}}, \bibinfo {author} {\bibfnamefont
  {G.~Q.}\ \bibnamefont {Zhao}}, \bibinfo {author} {\bibfnamefont {L.~Y.}\
  \bibnamefont {Xing}}, \bibinfo {author} {\bibfnamefont {G.}~\bibnamefont
  {Fabbris}}, \bibinfo {author} {\bibfnamefont {Y.~G.}\ \bibnamefont {Wang}},
  \bibinfo {author} {\bibfnamefont {Z.}~\bibnamefont {Deng}}, \bibinfo {author}
  {\bibfnamefont {Y.~J.}\ \bibnamefont {Uemura}}, \bibinfo {author}
  {\bibfnamefont {H.~K.}\ \bibnamefont {Mao}}, \bibinfo {author} {\bibfnamefont
  {D.}~\bibnamefont {Haskel}}, \bibinfo {author} {\bibfnamefont {W.~G.}\
  \bibnamefont {Yang}}, \ and\ \bibinfo {author} {\bibfnamefont {C.~Q.}\
  \bibnamefont {Jin}},\ }\href {\doibase 10.1103/PhysRevB.93.224403} {\bibfield
   {journal} {\bibinfo  {journal} {Phys. Rev. B}\ }\textbf {\bibinfo {volume}
  {93}},\ \bibinfo {pages} {224403} (\bibinfo {year} {2016})}\BibitemShut
  {NoStop}%
\bibitem [{\citenamefont {Glasbrenner}\ \emph {et~al.}(2014)\citenamefont
  {Glasbrenner}, \citenamefont {\ifmmode \check{Z}\else
  \v{Z}\fi{}uti\ifmmode~\acute{c}\else \'{c}\fi{}},\ and\ \citenamefont
  {Mazin}}]{Glasbrenner:2014aa}%
  \BibitemOpen
  \bibfield  {author} {\bibinfo {author} {\bibfnamefont {J.~K.}\ \bibnamefont
  {Glasbrenner}}, \bibinfo {author} {\bibfnamefont {I.}~\bibnamefont {\ifmmode
  \check{Z}\else \v{Z}\fi{}uti\ifmmode~\acute{c}\else \'{c}\fi{}}}, \ and\
  \bibinfo {author} {\bibfnamefont {I.~I.}\ \bibnamefont {Mazin}},\ }\href
  {\doibase 10.1103/PhysRevB.90.140403} {\bibfield  {journal} {\bibinfo
  {journal} {Phys. Rev. B}\ }\textbf {\bibinfo {volume} {90}},\ \bibinfo
  {pages} {140403} (\bibinfo {year} {2014})}\BibitemShut {NoStop}%
\bibitem [{\citenamefont {Yang}\ \emph {et~al.}(2015)\citenamefont {Yang},
  \citenamefont {Luo},\ and\ \citenamefont {Xiong}}]{Yang:2015aa}%
  \BibitemOpen
  \bibfield  {author} {\bibinfo {author} {\bibfnamefont {J.}~\bibnamefont
  {Yang}}, \bibinfo {author} {\bibfnamefont {S.}~\bibnamefont {Luo}}, \ and\
  \bibinfo {author} {\bibfnamefont {Y.}~\bibnamefont {Xiong}},\ }\href
  {\doibase http://dx.doi.org/10.1016/j.solidstatesciences.2015.06.007}
  {\bibfield  {journal} {\bibinfo  {journal} {Solid State Sci.}\ }\textbf
  {\bibinfo {volume} {46}},\ \bibinfo {pages} {102 } (\bibinfo {year}
  {2015})}\BibitemShut {NoStop}%
\bibitem [{\citenamefont {Wang}\ \emph {et~al.}(2017)\citenamefont {Wang},
  \citenamefont {Huang}, \citenamefont {Zhao}, \citenamefont {Yu},
  \citenamefont {Deng}, \citenamefont {Jin}, \citenamefont {Jia}, \citenamefont
  {Chen}, \citenamefont {Yang}, \citenamefont {Jiang},\ and\ \citenamefont
  {Cao}}]{Wang:2017aa}%
  \BibitemOpen
  \bibfield  {author} {\bibinfo {author} {\bibfnamefont {R.}~\bibnamefont
  {Wang}}, \bibinfo {author} {\bibfnamefont {Z.~X.}\ \bibnamefont {Huang}},
  \bibinfo {author} {\bibfnamefont {G.~Q.}\ \bibnamefont {Zhao}}, \bibinfo
  {author} {\bibfnamefont {S.}~\bibnamefont {Yu}}, \bibinfo {author}
  {\bibfnamefont {Z.}~\bibnamefont {Deng}}, \bibinfo {author} {\bibfnamefont
  {C.~Q.}\ \bibnamefont {Jin}}, \bibinfo {author} {\bibfnamefont {Q.~J.}\
  \bibnamefont {Jia}}, \bibinfo {author} {\bibfnamefont {Y.}~\bibnamefont
  {Chen}}, \bibinfo {author} {\bibfnamefont {T.~Y.}\ \bibnamefont {Yang}},
  \bibinfo {author} {\bibfnamefont {X.~M.}\ \bibnamefont {Jiang}}, \ and\
  \bibinfo {author} {\bibfnamefont {L.~X.}\ \bibnamefont {Cao}},\ }\href
  {\doibase 10.1063/1.4982713} {\bibfield  {journal} {\bibinfo  {journal} {AIP
  Advances}\ }\textbf {\bibinfo {volume} {7}},\ \bibinfo {pages} {045017}
  (\bibinfo {year} {2017})}\BibitemShut {NoStop}%
\bibitem [{\citenamefont {Abolfath}\ \emph {et~al.}(2001)\citenamefont
  {Abolfath}, \citenamefont {Jungwirth}, \citenamefont {Brum},\ and\
  \citenamefont {MacDonald}}]{Abolfath:2001aa}%
  \BibitemOpen
  \bibfield  {author} {\bibinfo {author} {\bibfnamefont {M.}~\bibnamefont
  {Abolfath}}, \bibinfo {author} {\bibfnamefont {T.}~\bibnamefont {Jungwirth}},
  \bibinfo {author} {\bibfnamefont {J.}~\bibnamefont {Brum}}, \ and\ \bibinfo
  {author} {\bibfnamefont {A.~H.}\ \bibnamefont {MacDonald}},\ }\href {\doibase
  10.1103/PhysRevB.63.054418} {\bibfield  {journal} {\bibinfo  {journal} {Phys.
  Rev. B}\ }\textbf {\bibinfo {volume} {63}},\ \bibinfo {pages} {054418}
  (\bibinfo {year} {2001})}\BibitemShut {NoStop}%
\bibitem [{\citenamefont {Dietl}\ \emph {et~al.}(2001)\citenamefont {Dietl},
  \citenamefont {Ohno},\ and\ \citenamefont {Matsukura}}]{Dietl:2001aa}%
  \BibitemOpen
  \bibfield  {author} {\bibinfo {author} {\bibfnamefont {T.}~\bibnamefont
  {Dietl}}, \bibinfo {author} {\bibfnamefont {H.}~\bibnamefont {Ohno}}, \ and\
  \bibinfo {author} {\bibfnamefont {F.}~\bibnamefont {Matsukura}},\ }\href
  {\doibase 10.1103/PhysRevB.63.195205} {\bibfield  {journal} {\bibinfo
  {journal} {Phys. Rev. B}\ }\textbf {\bibinfo {volume} {63}},\ \bibinfo
  {pages} {195205} (\bibinfo {year} {2001})}\BibitemShut {NoStop}%
\bibitem [{\citenamefont {Liu}\ \emph {et~al.}(2003)\citenamefont {Liu},
  \citenamefont {Sasaki},\ and\ \citenamefont {Furdyna}}]{Liu:2003aa}%
  \BibitemOpen
  \bibfield  {author} {\bibinfo {author} {\bibfnamefont {X.}~\bibnamefont
  {Liu}}, \bibinfo {author} {\bibfnamefont {Y.}~\bibnamefont {Sasaki}}, \ and\
  \bibinfo {author} {\bibfnamefont {J.~K.}\ \bibnamefont {Furdyna}},\ }\href
  {\doibase 10.1103/PhysRevB.67.205204} {\bibfield  {journal} {\bibinfo
  {journal} {Phys. Rev. B}\ }\textbf {\bibinfo {volume} {67}},\ \bibinfo
  {pages} {205204} (\bibinfo {year} {2003})}\BibitemShut {NoStop}%
\bibitem [{\citenamefont {Sawicki}\ \emph {et~al.}(2004)\citenamefont
  {Sawicki}, \citenamefont {Matsukura}, \citenamefont {Idziaszek},
  \citenamefont {Dietl}, \citenamefont {Schott}, \citenamefont {Ruester},
  \citenamefont {Gould}, \citenamefont {Karczewski}, \citenamefont {Schmidt},\
  and\ \citenamefont {Molenkamp}}]{Sawicki:2004aa}%
  \BibitemOpen
  \bibfield  {author} {\bibinfo {author} {\bibfnamefont {M.}~\bibnamefont
  {Sawicki}}, \bibinfo {author} {\bibfnamefont {F.}~\bibnamefont {Matsukura}},
  \bibinfo {author} {\bibfnamefont {A.}~\bibnamefont {Idziaszek}}, \bibinfo
  {author} {\bibfnamefont {T.}~\bibnamefont {Dietl}}, \bibinfo {author}
  {\bibfnamefont {G.~M.}\ \bibnamefont {Schott}}, \bibinfo {author}
  {\bibfnamefont {C.}~\bibnamefont {Ruester}}, \bibinfo {author} {\bibfnamefont
  {C.}~\bibnamefont {Gould}}, \bibinfo {author} {\bibfnamefont
  {G.}~\bibnamefont {Karczewski}}, \bibinfo {author} {\bibfnamefont
  {G.}~\bibnamefont {Schmidt}}, \ and\ \bibinfo {author} {\bibfnamefont
  {L.~W.}\ \bibnamefont {Molenkamp}},\ }\href {\doibase
  10.1103/PhysRevB.70.245325} {\bibfield  {journal} {\bibinfo  {journal} {Phys.
  Rev. B}\ }\textbf {\bibinfo {volume} {70}},\ \bibinfo {pages} {245325}
  (\bibinfo {year} {2004})}\BibitemShut {NoStop}%
\bibitem [{\citenamefont {Zemen}\ \emph {et~al.}(2009)\citenamefont {Zemen},
  \citenamefont {Ku{\v c}era}, \citenamefont {Olejn{\'\i}k},\ and\
  \citenamefont {Jungwirth}}]{Zemen:2009aa}%
  \BibitemOpen
  \bibfield  {author} {\bibinfo {author} {\bibfnamefont {J.}~\bibnamefont
  {Zemen}}, \bibinfo {author} {\bibfnamefont {J.}~\bibnamefont {Ku{\v c}era}},
  \bibinfo {author} {\bibfnamefont {K.}~\bibnamefont {Olejn{\'\i}k}}, \ and\
  \bibinfo {author} {\bibfnamefont {T.}~\bibnamefont {Jungwirth}},\ }\href
  {\doibase 10.1103/PhysRevB.80.155203} {\bibfield  {journal} {\bibinfo
  {journal} {Phys. Rev. B}\ }\textbf {\bibinfo {volume} {80}},\ \bibinfo
  {pages} {155203} (\bibinfo {year} {2009})}\BibitemShut {NoStop}%
\bibitem [{\citenamefont {van~der Laan}\ \emph {et~al.}(2010)\citenamefont
  {van~der Laan}, \citenamefont {Chopdekar}, \citenamefont {Suzuki},\ and\
  \citenamefont {Arenholz}}]{Laan:2010ab}%
  \BibitemOpen
  \bibfield  {author} {\bibinfo {author} {\bibfnamefont {G.}~\bibnamefont
  {van~der Laan}}, \bibinfo {author} {\bibfnamefont {R.~V.}\ \bibnamefont
  {Chopdekar}}, \bibinfo {author} {\bibfnamefont {Y.}~\bibnamefont {Suzuki}}, \
  and\ \bibinfo {author} {\bibfnamefont {E.}~\bibnamefont {Arenholz}},\ }\href
  {\doibase 10.1103/PhysRevLett.105.067405} {\bibfield  {journal} {\bibinfo
  {journal} {Phys. Rev. Lett.}\ }\textbf {\bibinfo {volume} {105}},\ \bibinfo
  {pages} {067405} (\bibinfo {year} {2010})}\BibitemShut {NoStop}%
\bibitem [{\citenamefont {Shibata}\ \emph {et~al.}(2018)\citenamefont
  {Shibata}, \citenamefont {Kitamura}, \citenamefont {Minohara}, \citenamefont
  {Yoshimatsu}, \citenamefont {Kadono}, \citenamefont {Ishigami}, \citenamefont
  {Harano}, \citenamefont {Takahashi}, \citenamefont {Sakamoto}, \citenamefont
  {Nonaka}, \citenamefont {Keisuke}, \citenamefont {Zhendong}, \citenamefont
  {Mitsuo}, \citenamefont {Shuichiro}, \citenamefont {Makoto}, \citenamefont
  {Jun-ichi}, \citenamefont {Akira}, \citenamefont {Kazunori}, \citenamefont
  {Hideyuki}, \citenamefont {Seiichi}, \citenamefont {Arata}, \citenamefont
  {Hiroshi}, \citenamefont {Tsuneharu},\ and\ \citenamefont
  {Atsushi}}]{Shibata:2018aa}%
  \BibitemOpen
  \bibfield  {author} {\bibinfo {author} {\bibfnamefont {G.}~\bibnamefont
  {Shibata}}, \bibinfo {author} {\bibfnamefont {M.}~\bibnamefont {Kitamura}},
  \bibinfo {author} {\bibfnamefont {M.}~\bibnamefont {Minohara}}, \bibinfo
  {author} {\bibfnamefont {K.}~\bibnamefont {Yoshimatsu}}, \bibinfo {author}
  {\bibfnamefont {T.}~\bibnamefont {Kadono}}, \bibinfo {author} {\bibfnamefont
  {K.}~\bibnamefont {Ishigami}}, \bibinfo {author} {\bibfnamefont
  {T.}~\bibnamefont {Harano}}, \bibinfo {author} {\bibfnamefont
  {Y.}~\bibnamefont {Takahashi}}, \bibinfo {author} {\bibfnamefont
  {S.}~\bibnamefont {Sakamoto}}, \bibinfo {author} {\bibfnamefont
  {Y.}~\bibnamefont {Nonaka}}, \bibinfo {author} {\bibfnamefont
  {I.}~\bibnamefont {Keisuke}}, \bibinfo {author} {\bibfnamefont
  {C.}~\bibnamefont {Zhendong}}, \bibinfo {author} {\bibfnamefont
  {F.}~\bibnamefont {Mitsuo}}, \bibinfo {author} {\bibfnamefont
  {F.}~\bibnamefont {Shuichiro}}, \bibinfo {author} {\bibfnamefont
  {O.}~\bibnamefont {Makoto}}, \bibinfo {author} {\bibfnamefont
  {F.}~\bibnamefont {Jun-ichi}}, \bibinfo {author} {\bibfnamefont
  {U.}~\bibnamefont {Akira}}, \bibinfo {author} {\bibfnamefont
  {W.}~\bibnamefont {Kazunori}}, \bibinfo {author} {\bibfnamefont
  {F.}~\bibnamefont {Hideyuki}}, \bibinfo {author} {\bibfnamefont
  {F.}~\bibnamefont {Seiichi}}, \bibinfo {author} {\bibfnamefont
  {T.}~\bibnamefont {Arata}}, \bibinfo {author} {\bibfnamefont
  {K.}~\bibnamefont {Hiroshi}}, \bibinfo {author} {\bibfnamefont
  {K.}~\bibnamefont {Tsuneharu}}, \ and\ \bibinfo {author} {\bibfnamefont
  {F.}~\bibnamefont {Atsushi}},\ }\href
  {http://dx.doi.org/10.1038/s41535-018-0077-4} {\bibfield  {journal} {\bibinfo
   {journal} {npj Quantum Materials}\ }\textbf {\bibinfo {volume} {3}}
  (\bibinfo {year} {2018})}\BibitemShut {NoStop}%
\bibitem [{\citenamefont {Mamiya}\ \emph {et~al.}(2006)\citenamefont {Mamiya},
  \citenamefont {Koide}, \citenamefont {Ishida}, \citenamefont {Osafune},
  \citenamefont {Fujimori}, \citenamefont {Suzuki}, \citenamefont {Katayama},\
  and\ \citenamefont {Yuasa}}]{Mamiya:2006ab}%
  \BibitemOpen
  \bibfield  {author} {\bibinfo {author} {\bibfnamefont {K.}~\bibnamefont
  {Mamiya}}, \bibinfo {author} {\bibfnamefont {T.}~\bibnamefont {Koide}},
  \bibinfo {author} {\bibfnamefont {Y.}~\bibnamefont {Ishida}}, \bibinfo
  {author} {\bibfnamefont {Y.}~\bibnamefont {Osafune}}, \bibinfo {author}
  {\bibfnamefont {A.}~\bibnamefont {Fujimori}}, \bibinfo {author}
  {\bibfnamefont {Y.}~\bibnamefont {Suzuki}}, \bibinfo {author} {\bibfnamefont
  {T.}~\bibnamefont {Katayama}}, \ and\ \bibinfo {author} {\bibfnamefont
  {S.}~\bibnamefont {Yuasa}},\ }\href {\doibase
  https://doi.org/10.1016/j.radphyschem.2005.07.042} {\bibfield  {journal}
  {\bibinfo  {journal} {Radiation Physics and Chemistry}\ }\textbf {\bibinfo
  {volume} {75}},\ \bibinfo {pages} {1872 } (\bibinfo {year} {2006})},\
  \bibinfo {note} {proceedings of the 20th International Conference on X-ray
  and Inner-Shell Processes}\BibitemShut {NoStop}%
\bibitem [{\citenamefont {Momma}\ and\ \citenamefont
  {Izumi}(2011)}]{Momma:2011aa}%
  \BibitemOpen
  \bibfield  {author} {\bibinfo {author} {\bibfnamefont {K.}~\bibnamefont
  {Momma}}\ and\ \bibinfo {author} {\bibfnamefont {F.}~\bibnamefont {Izumi}},\
  }\href {\doibase 10.1107/S0021889811038970} {\bibfield  {journal} {\bibinfo
  {journal} {J. Appl. Crystallogr.}\ }\textbf {\bibinfo {volume} {44}},\
  \bibinfo {pages} {1272} (\bibinfo {year} {2011})}\BibitemShut {NoStop}%
\bibitem [{Sup()}]{Suppl_BKZMA}%
  \BibitemOpen
  \href@noop {} {\bibinfo  {journal} {See supplemental material for the details
  of the sample growh and the cluster-model calculation. Calculations with
  various $D_{2d}$ splitting parameters are also shown}\ }\BibitemShut
  {NoStop}%
\bibitem [{\citenamefont {Furuse}\ \emph {et~al.}(2013)\citenamefont {Furuse},
  \citenamefont {Okano}, \citenamefont {Fuchino}, \citenamefont {Uchida},
  \citenamefont {Fujihira}, \citenamefont {Fujihira}, \citenamefont {Kadono},
  \citenamefont {Fujimori},\ and\ \citenamefont {Koide}}]{Furuse:2013aa}%
  \BibitemOpen
\bibfield  {journal} {  }\bibfield  {author} {\bibinfo {author} {\bibfnamefont
  {M.}~\bibnamefont {Furuse}}, \bibinfo {author} {\bibfnamefont
  {M.}~\bibnamefont {Okano}}, \bibinfo {author} {\bibfnamefont
  {S.}~\bibnamefont {Fuchino}}, \bibinfo {author} {\bibfnamefont
  {A.}~\bibnamefont {Uchida}}, \bibinfo {author} {\bibfnamefont
  {J.}~\bibnamefont {Fujihira}}, \bibinfo {author} {\bibfnamefont
  {S.}~\bibnamefont {Fujihira}}, \bibinfo {author} {\bibfnamefont
  {T.}~\bibnamefont {Kadono}}, \bibinfo {author} {\bibfnamefont
  {A.}~\bibnamefont {Fujimori}}, \ and\ \bibinfo {author} {\bibfnamefont
  {T.}~\bibnamefont {Koide}},\ }\href {\doibase 10.1109/TASC.2012.2236599}
  {\bibfield  {journal} {\bibinfo  {journal} {IEEE Trans. Appl. Supercond.}\
  }\textbf {\bibinfo {volume} {23}},\ \bibinfo {pages} {4100704} (\bibinfo
  {year} {2013})}\BibitemShut {NoStop}%
\bibitem [{\citenamefont {Nakajima}\ \emph {et~al.}(1999)\citenamefont
  {Nakajima}, \citenamefont {St\"ohr},\ and\ \citenamefont
  {Idzerda}}]{Nakajima:1999aa}%
  \BibitemOpen
  \bibfield  {author} {\bibinfo {author} {\bibfnamefont {R.}~\bibnamefont
  {Nakajima}}, \bibinfo {author} {\bibfnamefont {J.}~\bibnamefont {St\"ohr}}, \
  and\ \bibinfo {author} {\bibfnamefont {Y.~U.}\ \bibnamefont {Idzerda}},\
  }\href {\doibase 10.1103/PhysRevB.59.6421} {\bibfield  {journal} {\bibinfo
  {journal} {Phys. Rev. B}\ }\textbf {\bibinfo {volume} {59}},\ \bibinfo
  {pages} {6421} (\bibinfo {year} {1999})}\BibitemShut {NoStop}%
\bibitem [{\citenamefont {Tanaka}\ and\ \citenamefont
  {Jo}(1994)}]{Tanaka:1994aa}%
  \BibitemOpen
  \bibfield  {author} {\bibinfo {author} {\bibfnamefont {A.}~\bibnamefont
  {Tanaka}}\ and\ \bibinfo {author} {\bibfnamefont {T.}~\bibnamefont {Jo}},\
  }\href {\doibase 10.1143/JPSJ.63.2788} {\bibfield  {journal} {\bibinfo
  {journal} {Journal of the Physical Society of Japan}\ }\textbf {\bibinfo
  {volume} {63}},\ \bibinfo {pages} {2788} (\bibinfo {year}
  {1994})}\BibitemShut {NoStop}%
\bibitem [{\citenamefont {Kobayashi}\ \emph {et~al.}(2016)\citenamefont
  {Kobayashi}, \citenamefont {Ohya}, \citenamefont {Muneta}, \citenamefont
  {Takeda}, \citenamefont {Harada}, \citenamefont {Krempasky}, \citenamefont
  {Schmitt}, \citenamefont {Oshima}, \citenamefont {Strocov}, \citenamefont
  {Tanaka},\ and\ \citenamefont {Fujimori}}]{Kobayashi:2016aa}%
  \BibitemOpen
  \bibfield  {author} {\bibinfo {author} {\bibfnamefont {M.}~\bibnamefont
  {Kobayashi}}, \bibinfo {author} {\bibfnamefont {S.}~\bibnamefont {Ohya}},
  \bibinfo {author} {\bibfnamefont {I.}~\bibnamefont {Muneta}}, \bibinfo
  {author} {\bibfnamefont {Y.}~\bibnamefont {Takeda}}, \bibinfo {author}
  {\bibfnamefont {Y.}~\bibnamefont {Harada}}, \bibinfo {author} {\bibfnamefont
  {J.}~\bibnamefont {Krempasky}}, \bibinfo {author} {\bibfnamefont
  {T.}~\bibnamefont {Schmitt}}, \bibinfo {author} {\bibfnamefont
  {M.}~\bibnamefont {Oshima}}, \bibinfo {author} {\bibfnamefont {V.~N.}\
  \bibnamefont {Strocov}}, \bibinfo {author} {\bibfnamefont {M.}~\bibnamefont
  {Tanaka}}, \ and\ \bibinfo {author} {\bibfnamefont {A.}~\bibnamefont
  {Fujimori}},\ }\href {https://arxiv.org/abs/1608.07718} {\  (\bibinfo {year}
  {2016})},\ \Eprint {http://arxiv.org/abs/1608.07718} {1608.07718}
  \BibitemShut {NoStop}%
\bibitem [{\citenamefont {Gu}\ and\ \citenamefont {Maekawa}(2016)}]{Gu:2016aa}%
  \BibitemOpen
  \bibfield  {author} {\bibinfo {author} {\bibfnamefont {B.}~\bibnamefont
  {Gu}}\ and\ \bibinfo {author} {\bibfnamefont {S.}~\bibnamefont {Maekawa}},\
  }\href {\doibase 10.1103/PhysRevB.94.155202} {\bibfield  {journal} {\bibinfo
  {journal} {Phys. Rev. B}\ }\textbf {\bibinfo {volume} {94}},\ \bibinfo
  {pages} {155202} (\bibinfo {year} {2016})}\BibitemShut {NoStop}%
\bibitem [{\citenamefont {Takeda}\ \emph {et~al.}(2008)\citenamefont {Takeda},
  \citenamefont {Kobayashi}, \citenamefont {Okane}, \citenamefont {Ohkochi},
  \citenamefont {Okamoto}, \citenamefont {Saitoh}, \citenamefont {Kobayashi},
  \citenamefont {Yamagami}, \citenamefont {Fujimori}, \citenamefont {Tanaka},
  \citenamefont {Okabayashi}, \citenamefont {Oshima}, \citenamefont {Ohya},
  \citenamefont {Hai},\ and\ \citenamefont {Tanaka}}]{Takeda:2008aa}%
  \BibitemOpen
  \bibfield  {author} {\bibinfo {author} {\bibfnamefont {Y.}~\bibnamefont
  {Takeda}}, \bibinfo {author} {\bibfnamefont {M.}~\bibnamefont {Kobayashi}},
  \bibinfo {author} {\bibfnamefont {T.}~\bibnamefont {Okane}}, \bibinfo
  {author} {\bibfnamefont {T.}~\bibnamefont {Ohkochi}}, \bibinfo {author}
  {\bibfnamefont {J.}~\bibnamefont {Okamoto}}, \bibinfo {author} {\bibfnamefont
  {Y.}~\bibnamefont {Saitoh}}, \bibinfo {author} {\bibfnamefont
  {K.}~\bibnamefont {Kobayashi}}, \bibinfo {author} {\bibfnamefont
  {H.}~\bibnamefont {Yamagami}}, \bibinfo {author} {\bibfnamefont
  {A.}~\bibnamefont {Fujimori}}, \bibinfo {author} {\bibfnamefont
  {A.}~\bibnamefont {Tanaka}}, \bibinfo {author} {\bibfnamefont
  {J.}~\bibnamefont {Okabayashi}}, \bibinfo {author} {\bibfnamefont
  {M.}~\bibnamefont {Oshima}}, \bibinfo {author} {\bibfnamefont
  {S.}~\bibnamefont {Ohya}}, \bibinfo {author} {\bibfnamefont {P.~N.}\
  \bibnamefont {Hai}}, \ and\ \bibinfo {author} {\bibfnamefont
  {M.}~\bibnamefont {Tanaka}},\ }\href {\doibase
  10.1103/PhysRevLett.100.247202} {\bibfield  {journal} {\bibinfo  {journal}
  {Phys. Rev. Lett.}\ }\textbf {\bibinfo {volume} {100}},\ \bibinfo {pages}
  {247202} (\bibinfo {year} {2008})}\BibitemShut {NoStop}%
\bibitem [{\citenamefont {Edmonds}\ \emph {et~al.}(2005)\citenamefont
  {Edmonds}, \citenamefont {Farley}, \citenamefont {Johal}, \citenamefont
  {van~der Laan}, \citenamefont {Campion}, \citenamefont {Gallagher},\ and\
  \citenamefont {Foxon}}]{Edmonds:2005aa}%
  \BibitemOpen
  \bibfield  {author} {\bibinfo {author} {\bibfnamefont {K.~W.}\ \bibnamefont
  {Edmonds}}, \bibinfo {author} {\bibfnamefont {N.~R.~S.}\ \bibnamefont
  {Farley}}, \bibinfo {author} {\bibfnamefont {T.~K.}\ \bibnamefont {Johal}},
  \bibinfo {author} {\bibfnamefont {G.}~\bibnamefont {van~der Laan}}, \bibinfo
  {author} {\bibfnamefont {R.~P.}\ \bibnamefont {Campion}}, \bibinfo {author}
  {\bibfnamefont {B.~L.}\ \bibnamefont {Gallagher}}, \ and\ \bibinfo {author}
  {\bibfnamefont {C.~T.}\ \bibnamefont {Foxon}},\ }\href {\doibase
  10.1103/PhysRevB.71.064418} {\bibfield  {journal} {\bibinfo  {journal} {Phys.
  Rev. B}\ }\textbf {\bibinfo {volume} {71}},\ \bibinfo {pages} {064418}
  (\bibinfo {year} {2005})}\BibitemShut {NoStop}%
\bibitem [{\citenamefont {Thole}\ \emph {et~al.}(1992)\citenamefont {Thole},
  \citenamefont {Carra}, \citenamefont {Sette},\ and\ \citenamefont {van~der
  Laan}}]{Thole:1992aa}%
  \BibitemOpen
  \bibfield  {author} {\bibinfo {author} {\bibfnamefont {B.~T.}\ \bibnamefont
  {Thole}}, \bibinfo {author} {\bibfnamefont {P.}~\bibnamefont {Carra}},
  \bibinfo {author} {\bibfnamefont {F.}~\bibnamefont {Sette}}, \ and\ \bibinfo
  {author} {\bibfnamefont {G.}~\bibnamefont {van~der Laan}},\ }\href {\doibase
  10.1103/PhysRevLett.68.1943} {\bibfield  {journal} {\bibinfo  {journal}
  {Phys. Rev. Lett.}\ }\textbf {\bibinfo {volume} {68}},\ \bibinfo {pages}
  {1943} (\bibinfo {year} {1992})}\BibitemShut {NoStop}%
\bibitem [{\citenamefont {Carra}\ \emph {et~al.}(1993)\citenamefont {Carra},
  \citenamefont {Thole}, \citenamefont {Altarelli},\ and\ \citenamefont
  {Wang}}]{Carra:1993aa}%
  \BibitemOpen
  \bibfield  {author} {\bibinfo {author} {\bibfnamefont {P.}~\bibnamefont
  {Carra}}, \bibinfo {author} {\bibfnamefont {B.~T.}\ \bibnamefont {Thole}},
  \bibinfo {author} {\bibfnamefont {M.}~\bibnamefont {Altarelli}}, \ and\
  \bibinfo {author} {\bibfnamefont {X.}~\bibnamefont {Wang}},\ }\href {\doibase
  10.1103/PhysRevLett.70.694} {\bibfield  {journal} {\bibinfo  {journal} {Phys.
  Rev. Lett.}\ }\textbf {\bibinfo {volume} {70}},\ \bibinfo {pages} {694}
  (\bibinfo {year} {1993})}\BibitemShut {NoStop}%
\bibitem [{\citenamefont {Yang}\ \emph {et~al.}(2011)\citenamefont {Yang},
  \citenamefont {Chshiev}, \citenamefont {Dieny}, \citenamefont {Lee},
  \citenamefont {Manchon},\ and\ \citenamefont {Shin}}]{Yang:2011aa}%
  \BibitemOpen
  \bibfield  {author} {\bibinfo {author} {\bibfnamefont {H.~X.}\ \bibnamefont
  {Yang}}, \bibinfo {author} {\bibfnamefont {M.}~\bibnamefont {Chshiev}},
  \bibinfo {author} {\bibfnamefont {B.}~\bibnamefont {Dieny}}, \bibinfo
  {author} {\bibfnamefont {J.~H.}\ \bibnamefont {Lee}}, \bibinfo {author}
  {\bibfnamefont {A.}~\bibnamefont {Manchon}}, \ and\ \bibinfo {author}
  {\bibfnamefont {K.~H.}\ \bibnamefont {Shin}},\ }\href {\doibase
  10.1103/PhysRevB.84.054401} {\bibfield  {journal} {\bibinfo  {journal} {Phys.
  Rev. B}\ }\textbf {\bibinfo {volume} {84}},\ \bibinfo {pages} {054401}
  (\bibinfo {year} {2011})}\BibitemShut {NoStop}%
\bibitem [{\citenamefont {Okabayashi}\ \emph {et~al.}(2014)\citenamefont
  {Okabayashi}, \citenamefont {Koo}, \citenamefont {Sukegawa}, \citenamefont
  {Mitani}, \citenamefont {Takagi},\ and\ \citenamefont
  {Yokoyama}}]{Okabayashi:2014aa}%
  \BibitemOpen
  \bibfield  {author} {\bibinfo {author} {\bibfnamefont {J.}~\bibnamefont
  {Okabayashi}}, \bibinfo {author} {\bibfnamefont {J.~W.}\ \bibnamefont {Koo}},
  \bibinfo {author} {\bibfnamefont {H.}~\bibnamefont {Sukegawa}}, \bibinfo
  {author} {\bibfnamefont {S.}~\bibnamefont {Mitani}}, \bibinfo {author}
  {\bibfnamefont {Y.}~\bibnamefont {Takagi}}, \ and\ \bibinfo {author}
  {\bibfnamefont {T.}~\bibnamefont {Yokoyama}},\ }\href {\doibase
  10.1063/1.4896290} {\bibfield  {journal} {\bibinfo  {journal} {Applied
  Physics Letters}\ }\textbf {\bibinfo {volume} {105}},\ \bibinfo {pages}
  {122408} (\bibinfo {year} {2014})}\BibitemShut {NoStop}%
\bibitem [{\citenamefont {Nakajima}\ \emph {et~al.}(1998)\citenamefont
  {Nakajima}, \citenamefont {Koide}, \citenamefont {Shidara}, \citenamefont
  {Miyauchi}, \citenamefont {Fukutani}, \citenamefont {Fujimori}, \citenamefont
  {Iio}, \citenamefont {Katayama}, \citenamefont {N\'yvlt},\ and\ \citenamefont
  {Suzuki}}]{Nakajima:1998aa}%
  \BibitemOpen
  \bibfield  {author} {\bibinfo {author} {\bibfnamefont {N.}~\bibnamefont
  {Nakajima}}, \bibinfo {author} {\bibfnamefont {T.}~\bibnamefont {Koide}},
  \bibinfo {author} {\bibfnamefont {T.}~\bibnamefont {Shidara}}, \bibinfo
  {author} {\bibfnamefont {H.}~\bibnamefont {Miyauchi}}, \bibinfo {author}
  {\bibfnamefont {H.}~\bibnamefont {Fukutani}}, \bibinfo {author}
  {\bibfnamefont {A.}~\bibnamefont {Fujimori}}, \bibinfo {author}
  {\bibfnamefont {K.}~\bibnamefont {Iio}}, \bibinfo {author} {\bibfnamefont
  {T.}~\bibnamefont {Katayama}}, \bibinfo {author} {\bibfnamefont
  {M.}~\bibnamefont {N\'yvlt}}, \ and\ \bibinfo {author} {\bibfnamefont
  {Y.}~\bibnamefont {Suzuki}},\ }\href {\doibase 10.1103/PhysRevLett.81.5229}
  {\bibfield  {journal} {\bibinfo  {journal} {Phys. Rev. Lett.}\ }\textbf
  {\bibinfo {volume} {81}},\ \bibinfo {pages} {5229} (\bibinfo {year}
  {1998})}\BibitemShut {NoStop}%
\bibitem [{\citenamefont {Freeman}\ \emph {et~al.}(2006)\citenamefont
  {Freeman}, \citenamefont {Edmonds}, \citenamefont {van~der Laan},
  \citenamefont {Farley}, \citenamefont {Johal}, \citenamefont {Arenholz},
  \citenamefont {Campion}, \citenamefont {Foxon},\ and\ \citenamefont
  {Gallagher}}]{Freeman:2006aa}%
  \BibitemOpen
  \bibfield  {author} {\bibinfo {author} {\bibfnamefont {A.~A.}\ \bibnamefont
  {Freeman}}, \bibinfo {author} {\bibfnamefont {K.~W.}\ \bibnamefont
  {Edmonds}}, \bibinfo {author} {\bibfnamefont {G.}~\bibnamefont {van~der
  Laan}}, \bibinfo {author} {\bibfnamefont {N.~R.~S.}\ \bibnamefont {Farley}},
  \bibinfo {author} {\bibfnamefont {T.~K.}\ \bibnamefont {Johal}}, \bibinfo
  {author} {\bibfnamefont {E.}~\bibnamefont {Arenholz}}, \bibinfo {author}
  {\bibfnamefont {R.~P.}\ \bibnamefont {Campion}}, \bibinfo {author}
  {\bibfnamefont {C.~T.}\ \bibnamefont {Foxon}}, \ and\ \bibinfo {author}
  {\bibfnamefont {B.~L.}\ \bibnamefont {Gallagher}},\ }\href {\doibase
  10.1103/PhysRevB.73.233303} {\bibfield  {journal} {\bibinfo  {journal} {Phys.
  Rev. B}\ }\textbf {\bibinfo {volume} {73}},\ \bibinfo {pages} {233303}
  (\bibinfo {year} {2006})}\BibitemShut {NoStop}%
\end{thebibliography}%

\end{document}